\documentclass[a4paper,twoside]{article}

\usepackage{epsfig}
\usepackage{calc}
\usepackage{amssymb}
\usepackage{amstext}
\usepackage{amsmath}
\usepackage{multicol}
\usepackage{pslatex}
\usepackage[round]{natbib}
\usepackage{algorithm2e}
\usepackage[bottom]{footmisc}
\usepackage{booktabs}
\usepackage{multirow}
\usepackage{graphicx}
\usepackage{xcolor}
\usepackage[hidelinks]{hyperref}
\usepackage{SCITEPRESS}

\begin{document}

\title{Measuring Duplication, Leakage, and Rendering Dependence in Malware-as-Image Benchmarks}

\author{\authorname{Stefan Larson and Kevin Leach}
\affiliation{Vanderbilt University, Nashville, TN, USA}
\email{\{stefan.larson, kevin.leach\}@vanderbilt.edu}
}
\newcommand{\codeurl}{\url{https://github.com/gxlarson/malware-image-audit}}


\abstract{%
Classifying malware binaries rendered as images is a popular approach to malware family identification, and papers on the standard benchmarks for this task report accuracies of 97--99\%.
In this paper we audit ten of these benchmarks for near-duplicate images and test-train leakage, and retrain standard classifiers under naive and content-aware splits.
Deduplicating by file hash, which is the standard practice, does not prevent leakage, because distinct binaries render to identical images.
Deduplicating on image content reduces top-1 accuracy by up to 8 points and macro-F1 by up to 15 points on the released benchmarks we find leaky.
The rendering step also adds duplication, since byteplots of the same binaries contain 19.7\% near-duplicates where static feature vectors contain 2.2\%, and changing the width at which bytes are wrapped into rows reduces accuracy by 20 to 33 points, although no byte changes.
We recommend that content-level duplication and a non-image baseline be reported alongside accuracy.}

\onecolumn \maketitle \normalsize \setcounter{footnote}{0} \vfill

\section{\uppercase{Introduction}}
\label{sec:intro}

Representing malicious code (malware) as an image is a popular approach to malware family classification.
A binary is the file of raw bytes that contains a compiled program, and it can be read as a stream of unsigned 8-bit values and wrapped into rows of fixed width, which produces a grayscale image known as a \emph{byteplot}.
These images are used for \emph{family classification}: a malware \emph{family} is a group of programs derived from one code base, released as many variants by recompiling it, reconfiguring it, or running it through an obfuscator (a tool that rewrites a program so that its code differs but its behavior does not).

\citet{nataraj2011} observed that programs in the same family produce visually similar textures under the byteplot encoding, and reported 98.08\% accuracy across 25 families using GIST descriptors and a $k$-NN classifier.
A large body of work has followed, and several benchmarks built to support it are now standard, including MaleVis \citep{malevis}, Dumpware10 \citep{dumpware10}, Virus-MNIST \citep{virusmnist}, MalNet-Image \citep{malnet}, AndroDex \citep{androdex}, MalVis \citep{malvis} and MaleX \citep{mohammedmalware}.
Papers that train CNN and Vision Transformer models on these benchmarks routinely report accuracies in the 97--99\% range, and a recent survey of more than 100 papers in the area \citep{brosolo2025survey} and a systematization of the same literature \citep{brosolo2024sok} both report this range.

Despite this volume of work, the quality of these benchmarks has received little scrutiny.
Duplication is one concern, since the programs in a malware family are variants of a common program and their shared code produces identical or near-identical images.
The rendering is a second concern, since the width at which bytes are wrapped into rows is an arbitrary choice by the benchmark creators, and whether a model's accuracy depends on that choice has not been tested.

Leakage through duplicates is documented across machine-learning-based science \citep{kapoor2023leakage}, and it is present even in canonical vision benchmarks, where 3.3\% of CIFAR-10 and 10\% of CIFAR-100 test images have a near-duplicate in training \citep{barz2020train}.
The same defect is measured in document classification \citep{larson2023rvlcdip, larson2026revising}, face recognition \citep{wu2024identityoverlap}, license plate recognition \citep{laroca2023nearduplicates} and open-domain question answering \citep{lewis2021overlap}, and most of the studies that report accuracy find that models score lower on test data that does not have a counterpart in training.
Studies of Android malware datasets of extracted features report mixed results on both effects \citep{tesseract, zhao2021dedup}, and whether either holds for the malware-as-image benchmarks has not been measured.

In this paper, we measure duplicate content and test-train overlap in ten malware-as-image benchmarks and quantify the effect of both on reported accuracy.
We identify duplicates by the decoded pixels of each image, because benchmark creators deduplicate by the hash of the program file, and two different files can render to the same image.
On MalNet-Tiny, whose authors deduplicated by file hash, 14.9\% of the test images are pixel-identical to a training image.
We then build deduplicated, family-grouped splits and retrain standard classifiers on them, and we examine the rendering step itself, both as a source of duplication that the underlying corpus does not contain and as a choice on which the trained model turns out to depend.
We make our audit code, deduplicated splits and per-run results publicly available at \codeurl.

This paper is a benchmark audit, and we do not propose a new model, training procedure or dataset.
Our contribution is to measure data quality inside the image benchmarks themselves and to quantify its effect on reported accuracy.
This type of paper is in the tradition of prior work that has revealed similar problems in computer vision \citep{barz2020train, radenovic2018revisiting, northcutt2021labelerrors}, NLP \citep{lewis2021overlap} and document understanding \citep{larson2023rvlcdip, larson2026revising} benchmarks.

To summarize, the main contributions of this paper are:
\begin{enumerate}\itemsep2pt
\item We audit ten malware-as-image benchmarks, finding duplication rates from 0 to 35.7\% and test-set leakage as high as 50.6\%, and show that deduplicating by file hash, which is the standard practice, does not remove the leakage, since distinct files can contain identical code and render to identical images.
On MalNet-Tiny, file-hash deduplication does not remove any images, so it does not recover any of the 8.44-point drop in accuracy that we measure under content-aware deduplication.
\item We construct content-aware, family-grouped splits and retrain three architectures on the seven benchmarks small enough for us to train on repeatedly, and find that accuracy drops by 3.7 to 10.5 percentage points on five of them, while the drop on the other two is not significant.
A training set enlarged with further copies of its own images is no more accurate, so the drop comes from test images that have a copy in training and not from the smaller training set.
\item We measure the contribution of the image encoding itself on the two corpora where this measurement is possible.
MOTIF provides gold labels, raw binaries and static features together.
MOTIF's feature vectors contain 2.2\% near-duplicates and do not lose accuracy under deduplication, while our byteplots of the same samples contain 19.7\% and lose $10.52\pm2.23$ points.
Dumpware10 provides the same binaries at four byteplot widths, and training at one width and testing at another reduces accuracy by $20.53\pm2.71$ points, although the bytes do not change, and by $33.08\pm1.74$ points on our renderings of MOTIF at five widths.
\end{enumerate}

We do not claim that malware-as-image accuracy is uniformly an artifact of duplicated data, since Malimg loses less than one point under our deduplicated split.
Our claim is that inflation must be measured per benchmark, and that the image encoding is not a neutral re-representation of the binary.

\section{\uppercase{Background and Related Work}}
\label{sec:related}

\paragraph{Task setup.}
Malware family classification is posed as $n$-way image classification.
Each sample is one program rendered as one image, and its label is the family to which the program is assigned.
A model is trained on a training split and evaluated by top-1 accuracy on a held-out test split.
The benchmarks differ in which part of the program is rendered.
Most are rendered from the Windows PE (Portable Executable) file on disk, Dumpware10 from the memory that a program occupies while it runs, Virus-MNIST from only the first kilobyte of the file, and the Android benchmarks from the DEX (Dalvik Executable) bytecode inside the Android application package (APK), which is not native machine code.
Table~\ref{tab:datasets} lists the ten that we audit, with the part of the program from which each is rendered, its encoding and the source of its labels.
The task column is the label that we train and evaluate on: the malware family, the malware type (for example adware or trojan), a broader category, a $k$-means cluster, or benign versus malicious (detection).
``Single AV'' in the label column means the verdict of one antivirus vendor, while MOTIF's labels come from the threat reports of 14 vendors and are alias-resolved by the MOTIF authors, so several reported family names map to one label.
AndroDex contains each program twice, before and after obfuscation, and except where we state otherwise our AndroDex measurements use the 10{,}281 originals and hold out the obfuscated variants.
\begin{table*}[t]
\centering
\caption{The ten malware datasets analyzed in this paper.}
\label{tab:datasets}
\small
\setlength{\tabcolsep}{3pt}
\resizebox{\textwidth}{!}{%
\begin{tabular}{lllllll}
\toprule
\textbf{Dataset} & \textbf{Rendered from} & \textbf{Encoding} & \textbf{Images} & \textbf{Task} & \textbf{Classes} & \textbf{Label source} \\
\midrule
Malimg \citep{nataraj2011}     & whole PE file  & byte $\to$ gray          & 9{,}347       & family    & 25 families         & single AV \\
MaleVis \citep{malevis}        & whole PE file  & 3 bytes $\to$ RGB        & 14{,}226      & family    & 26 classes          & single AV (Comodo) \\
Virus-MNIST \citep{virusmnist} & first 1\,KB    & byte $\to$ gray, $32^2$  & 51{,}880 & cluster & 10 classes         & $k$-means pseudo \\
Dumpware10 \citep{dumpware10}  & memory dump    & 3 bytes $\to$ RGB        & 4{,}294       & family    & 11 classes          & single AV \\
MalNet-Tiny \citep{malnet}     & Android DEX    & byte $\to$ gray          & 87{,}430      & type      & 43 types            & Euphony/VirusTotal \\
MalNet-Image \citep{malnet}    & Android DEX    & byte $\to$ gray          & 1{,}262{,}025 & type, family & 47 types, 696 fam.\ & Euphony/VirusTotal \\
MalVis \citep{malvis}          & Android DEX    & entropy/$n$-gram   & 1{,}300{,}822 & category  & 11 classes          & AndroZoo aggregate \\
AndroDex \citep{androdex}      & Android DEX    & byte $\to$ gray          & 21{,}507 & detection & 2 (benign/mal.)    & dataset construction \\
MOTIF \citep{motif}            & whole PE file  & byte $\to$ gray          & 3{,}095       & family    & 454 families  & threat reports \\
MaleX \citep{mohammedmalware}            & whole PE file  & byte $\to$ gray, $256^2$ & 1{,}044{,}394 & detection, family & 2; 100 families & Avast, BitDefender \\
\bottomrule
\end{tabular}}
\end{table*}

\paragraph{File hashes and image content.}
Malware corpora are conventionally deduplicated by hashing the distributed file.
Consider two Android applications that differ only in a signing certificate and a version string.
Their APK files have different SHA-256 hashes, so deduplication by file hash keeps both, but their bytecode is identical and both render to the same image.
If one copy is in the training split and the other in the test split, the model is evaluated on an image that it has already seen in training.
We therefore distinguish an image's \emph{file hash}, over the file as distributed, from its \emph{content hash}, the SHA-256 of the decoded pixel array on which the model is trained.

\paragraph{Malware as images.}
\citet{nataraj2011} introduced the byteplot and reported 98.08\% accuracy with GIST features and $k$-NN.
They also noted that when variants of one family are processed by the same \emph{packer} (a tool that compresses a program and wraps it in a stub that restores it at run time), ``the images of the newly packed malware (of same family) are also similar,'' which is one source of the near-duplicates that we measure.
Later authors replaced GIST with CNNs and transformers and extended the idea to Android DEX bytecode and memory dumps \citep{malevis, dumpware10, malnet, malvis, androdex, mohammedmalware}, reporting accuracies of 97--99\% on their benchmarks.
Two recent surveys organize this literature by model and rendering scheme \citep{brosolo2025survey, brosolo2024sok}, but neither examines duplication or leakage.

\paragraph{Duplication in code and security datasets.}
There is a specific reason to expect duplication in malware datasets, since their samples are code and their classes are defined by shared code.
\citet{allamanis2019duplication} shows that duplicated code inflates the measured performance of models of code, and \citet{croft2023vulndata} find data quality problems in vulnerability datasets that affect reported detection performance.
TESSERACT \citep{tesseract} shows that ignoring temporal and spatial constraints inflates Android malware detection results.
Closest to our own measurement, \citet{zhao2021dedup} deduplicate Android datasets of extracted features and find that the accuracy of supervised classification is largely unchanged.
We find the same for the feature vectors of MOTIF, which do not show a drop under deduplication, whereas a byteplot rendering of the same samples loses ten points (Section~\ref{sec:repr}), so the two results differ in the representation and not in the corpus.
\citet{zhao2021dedup} deduplicate extracted features, whereas the image benchmarks are deduplicated, if at all, by file hash, and we identify duplicates in the decoded images.
Prior work does not measure duplication and leakage inside the image datasets themselves.

\section{\uppercase{Auditing the Benchmarks}}
\label{sec:audit}
We first set out how we index and compare images, then report what we find in the audit of the ten benchmarks.

\subsection{Method}
We audit the ten benchmarks listed in Table~\ref{tab:datasets}.
Nine are distributed as image files, which we decode into the pixel array on which a model is trained.
MOTIF is distributed as binaries and feature vectors, so we render its binaries as byteplots ourselves, using the convention described below.
MALITE \citep{malite} also renders MOTIF's binaries as grayscale byteplots and trains vision models on them, alongside Malimg, Dumpware10 and MalNet, so MOTIF is already used for image-based classification.
Of MOTIF's 454 malware families, 131 contain a single sample, and we train on the remaining 323 families, 2{,}964 samples in all.
All of the benchmarks are byteplots, with MalVis being the exception, which encodes entropy over a 32-byte window (red and blue channels) and $n$-gram frequencies (green) into a fixed $256\times256$ RGB image, and we include MalVis because it allows the duplication question to be asked of a different encoding.
MaleX distributes each of its programs both as a byteplot and as a bigram-DCT image, and we audit its byteplots and compare the two encodings in Section~\ref{sec:repr}.

A byteplot's width is a fixed constant chosen by the benchmark's creator, and its height is then determined by the file size, which means a larger program produces a taller image.
We measure the effect of this choice of width on accuracy in Section~\ref{sec:repr}.
Because image heights therefore vary within a benchmark while CNNs expect a fixed input, every image is resized to $224\times224$ before training, without preserving aspect ratio, as the creators of MaleVis and MalNet-Image also do \citep{malevis, malnet}.
The resizing discards the file-size information encoded in the image height and rescales the texture by a per-image factor, which is relevant to the cross-rendering result in Section~\ref{sec:repr}.

For each image we record its content hash and file hash (Section~\ref{sec:related}), two perceptual hashes (pHash and dHash, which give visually similar images similar hashes) and its dimensions.
Two images are \emph{exact content duplicates} when their content hashes match.
Hashing the decoded pixels also makes the measure independent of re-encoding, so mirrors of a benchmark that differ only in compression yield the same content hashes.

\begin{figure}[t]
\centering
\includegraphics[width=0.9\linewidth]{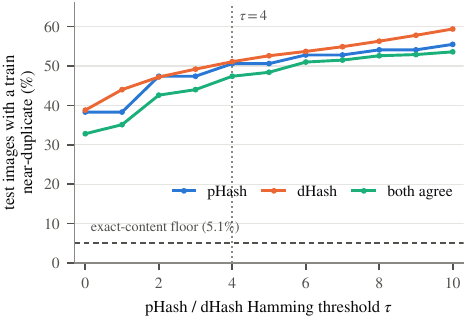}
\caption{Leakage in the MaleVis published split against the pHash threshold $\tau$.}
\label{fig:threshold}
\end{figure}

Two images are \emph{near-duplicates} when the Hamming distance between their 64-bit pHashes is at most a threshold $\tau$, which we set to 4.
Below we check that this threshold matches copies of one program and that our conclusions do not depend on $\tau$.
Figure~\ref{fig:threshold} shows the leakage in the MaleVis published split for $\tau$ from 0 to 10, which is 38.3\% at $\tau=0$, where only hash-identical images are matched, 50.6\% at $\tau=4$, and at least 38\% across the whole range.
At $\tau=0$, which merges only images with identical pHashes, the drop in accuracy on MaleVis is 3.67 percentage points (Section~\ref{sec:inflation}), so the drop does not depend on admitting near matches, and $\tau$ sets how much of it we measure.

We checked in two ways that $\tau=4$ groups copies of one program and does not group different programs.
First, if two images within distance 4 have different labels, the pair is either a false match or a labeling disagreement, so the share of such pairs is an upper bound on the rate of false matches.
On five of the seven benchmarks that we retrain on (Section~\ref{sec:inflation}) this share is at most 0.15\%, against 50--99\% for random pairs.
On Malimg and MalNet-Tiny it is 18.7\% and 10.5\%, but it is already 20.1\% and 7.7\% at $\tau=0$, so these pairs come from the labels and do not increase with the threshold: the Malimg pairs are all in one group of 106 \texttt{Autorun.K} and 800 \texttt{Yuner.A} images, and the largest mixed groups in MalNet-Tiny contain images with overlapping type labels such as adware and spyware.
Second, on MOTIF we can compare the binaries behind each pair of images, using the distance between their TLSH digests \citep{tlsh}, a locality-sensitive hash of a file's bytes for which a lower distance means more similar files.
The median TLSH distance is 1 for image pairs at pHash distance 0 and 4 for pairs at distance 1 to 4, against 280 for other pairs from the same family and 390 for pairs from different families.

We then merge near-duplicate pairs transitively (union-find over content-distinct representatives) into \emph{near-duplicate groups}, each containing one distinct image together with every copy of that image.
From here on we work with these groups, reducing each group to a single representative when building deduplicated splits and counting a test image as leaked when its group also contains a training image.
Where a benchmark provides a train/test split we measure leakage on that split, and where it does not, we use the same stratified random split as the papers that report results on it, and measure leakage there.

\begin{figure}[t]
\centering
\includegraphics[width=0.9\linewidth]{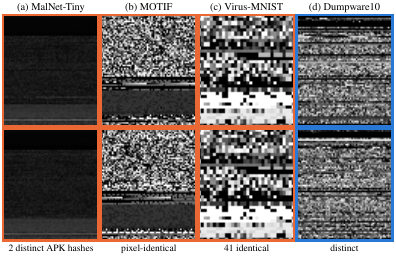}
\caption{Duplicate pairs found by the audit, one benchmark per column.
Orange frames mark duplicates and blue the negative control.}
\label{fig:gallery}
\end{figure}

Figure~\ref{fig:gallery} shows one pair of distinct entries per benchmark: (a) two MalNet-Tiny APKs with different SHA-256 hashes, identical pixels and opposite sides of the published split, (b) an exact duplicate in our rendering of MOTIF, which is deduplicated by construction, (c) one of 41 identical Virus-MNIST images, and (d) a Dumpware10 pair that the procedure reports as distinct, as a negative control.

\begin{table}[t]
\centering
\caption{Duplication audit results.
Test leakage is exact / including near-duplicates, n/m is not measured and --- is not applicable.
$^{\ddagger}$Hamming distance 0, so a lower bound.
$^{\P}$Cross-validated, so there is no single split.}
\label{tab:scorecard}
\small
\setlength{\tabcolsep}{3pt}
\resizebox{\linewidth}{!}{%
\begin{tabular}{llrrl}
\toprule
\textbf{Dataset} & \textbf{Split} & \textbf{Exact} & \textbf{Near} & \textbf{Test leakage} \\
\midrule
Malimg           & random    & 15.1\% & 22.5\% & 14.6\% / 33\% \\
MaleVis          & published & 7.3\%  & 59.5\% & 5.1\% / 50.6\% \\
Virus-MNIST      & ---       & 35.7\% & 11.0\% & --- \\
Dumpware10       & published & 0\%    & 22\%   & 0\% / 14.6\% \\
MalNet-Tiny      & published & 16.9\% & 26.4\% & 14.9\% / 36.2\% \\
MalNet-Image     & published & 16.6\% & 36.8\%$^{\ddagger}$ & 14.4\% / 40.7\% \\
MalVis           & random    & 15.6\% & 50.8\%$^{\ddagger}$ & 14.2\% / 48.3\%$^{\ddagger}$ \\
AndroDex (all)   & ---       & 18.5\% & 16.3\% & --- \\
AndroDex (originals) & random    & 35.7\% & 19.3\% & 32.3\% / 44.3\% \\
MOTIF            & 5-fold CV & 2.2\%  & 19.7\% & n/m$^{\P}$ \\
MaleX            & random    & 13.3\% & 32.9\% & 13.1\% / 40.5\% \\
\bottomrule
\end{tabular}}
\end{table}

\subsection{Duplication Audit Results}

Table~\ref{tab:scorecard} reports the audit over the ten benchmarks described in Table~\ref{tab:datasets}.
The Exact column is the share of all images that have at least one pixel-identical copy, and the Near column is the share of content-distinct images (one per exact group) that have at least one other distinct image within pHash distance 4, so Near does not count exact copies and can be smaller than Exact.
Rows marked published show the leakage in the benchmark's own partition, which applies to any paper that uses that partition, and rows marked random show it for a stratified random 80/20 split of the kind the literature on that benchmark constructs, averaged over three seeds.
For MalNet-Image we report its type split, and its family and binary splits have leakage within 0.1 points of it.
Virus-MNIST provides per-binary hashes only for its test split, so we cannot measure its leakage.
For comparison, leakage in the leaky benchmarks here is three to fifteen times the 3.3\% and 10\% near-duplicate rates measured in CIFAR-10 and CIFAR-100 \citep{barz2020train}, while the cleanest, Dumpware10 at 14.6\%, is within five points of CIFAR-100.

We report three findings:

\paragraph{(1) Deduplication by file hash.}
The authors of MalNet-Image report removing any sample whose APK file hash had already been collected, and we confirm this practice in our audit, finding 0\% file-hash reuse and no image whose file hash appears on both sides of the split.
Duplication in the pixel content remains, however, and in MalNet-Tiny, the subset small enough to retrain on repeatedly, 16.9\% of images are exact content duplicates and 14.9\% of the test set is pixel-identical to training data, and the full 1.26M-image set has a similar rate of 16.6\%.

MalVis shows the same gap between file hashes and image content independently, at 1.3M images and under a different encoding.
Each file hash occurs once and under a single family label, yet 15.6\% of its images are content duplicates and 14.2\% of a naive test split is pixel-identical to training data.
Because MalVis is not a byteplot, it is the stronger corroboration, since the duplication remains under a rendering that discards byte order and encodes entropy and $n$-gram statistics.
MaleX shows the same gap for Windows PE files, over 1.04M programs: each SHA-256 occurs once, yet 13.3\% of its byteplots have a pixel-identical copy, and 17{,}317 of them are in groups of pixel-identical byteplots that contain both malware and benign programs.

This gap arises because distinct APKs can contain identical DEX bytecode, as with repackaged applications, shared libraries and rebuilds that change only signing or manifest metadata, and identical bytecode renders to an identical image, so a hash of the distributed file does not identify the model's input.

Because MalNet-Tiny does not contain any file-hash reuse, deduplicating it by file hash does not remove any images and leaves the naive split intact, so a file-hash pipeline does not recover any of the drop in accuracy that we measure under content-aware deduplication (Section~\ref{sec:inflation}).
AndroDex is the converse case, over its 10{,}281 originals with the obfuscated variants set aside: file and content hashing remove the same 26.0\% of the corpus, while near-duplicate grouping removes 34.4\% (these are removal rates, the share of images removed when each group is reduced to one image, and they are lower than the rates in Table~\ref{tab:scorecard}, which count every image in a group), and of the 44.3\% of test images that have a duplicate in training in its naive split, file-hash deduplication removes only the 32.3 points that are pixel-identical.
Between them, the two cases show that neither identity hashing nor exact-content hashing is sufficient.

\paragraph{(2) Duplication by rendering scheme.}
The choice of which bytes to render accounts for the entire range of duplication rates in Table~\ref{tab:scorecard}, from 0 to 35.7\%.
Memory dumps (Dumpware10) do not contain any exact duplicates, since runtime state differs even between samples of one family; whole-file byteplots have rates of 7--15\% as shared headers and packer stubs recur; Android DEX renderings have rates of roughly 17\%; and truncated-byte images are worst, since Virus-MNIST renders only the first 1\,KB and shared prologues produce identical $32\times32$ images.
MalVis, which is not a byteplot, is in the middle of that range, so the effect is not specific to byte-to-pixel mapping.

\paragraph{(3) Leakage in published splits.}
Under the published train/test partitions, 50.6\% of the MaleVis test images, 36.2\% of the MalNet-Tiny test images and 40.7\% of the MalNet-Image test images have an exact or near-duplicate in training, meaning that published work using these partitions is evaluated with this leakage.
Malimg does not have an official split, and in the random partitions used by papers on Malimg roughly 33\% of the test images have a near-duplicate in training.

\paragraph{Label quality.}
We also find problems with labels in our audit, although duplication is the issue that we quantify end-to-end.
Virus-MNIST family labels are unsupervised $k$-means pseudo-labels, and re-running $k$-means on raw bytes recovers 77\% of the published labels (ARI 0.69), which means that reported accuracy is substantially circular.
Malimg contains degenerate classes such as \texttt{Obfuscator.AD}, which is 142 images forming a single near-duplicate cluster, and \texttt{Autorun.K}, which has the same structure.
MaleVis has an \texttt{Other} class that is not a family.
The 100 families of MaleX are Avast detection names: five families each appear as two classes that differ only in a suffix (\texttt{Win32:Allaple} and \texttt{Win32:Allaple [Wrm]}), 17 classes are generic detections such as \texttt{Win32:Trojan-gen}, and 19 programs are listed both as malware and as benign.

\section{\uppercase{Effect of Deduplication on Accuracy}}
\label{sec:inflation}

The audit in Section~\ref{sec:audit} measures how much duplicated content each benchmark contains.
In this section we retrain standard classifiers on splits in which no near-duplicate group spans training and test, and measure how much of the reported accuracy the duplication accounts for.

\subsection{Protocol}
We compare three splits per dataset, holding the training regime fixed:
\begin{description}\itemsep1pt
\item[naive] stratified random 80/20, applied uniformly so that the three conditions are comparable across benchmarks.
The published ratios differ (MaleVis provides a fixed 9{,}100/5{,}126 partition, MalNet-Tiny a 70/10/20 one and Dumpware10 an 80/20 one), so this condition represents the standard random split and is not the same as each benchmark's own split.
We also train and evaluate on the published splits of MaleVis, MalNet-Tiny and Dumpware10 (Section~\ref{sec:results}).
\item[dedup] near-duplicate groups (exact content $\cup$ pHash $\le4$) collapsed to one representative, then partitioned per family so that no near-duplicate spans the split.
\item[ctrl] a control for training-set size, evaluated on the \textbf{dedup} test set.
Its training set has the same size as that of \textbf{dedup} but is drawn at random from all images outside the \textbf{dedup} test set, so unlike \textbf{dedup} it can contain duplicates of other training images and of test images.
\end{description}
Deduplication reduces the size of the training set, and MaleVis---the worst case---loses 58\% of its images and is left with 4{,}757 for training, and Dumpware10's deduplicated training set is the smallest at 2{,}813.

Fine-tuning an ImageNet-pretrained network is the standard pipeline in this literature, as in the 97.48\% that MaleVis's authors report with DenseNet-121 \citep{malevis} and in the surveys of the area \citep{brosolo2025survey, brosolo2024sok}.
We therefore fine-tune ImageNet-pretrained ResNet-50 \citep{resnet}, DenseNet-121 \citep{densenet} and ViT-B/16 \citep{vit}, chosen to span a residual CNN, a densely connected CNN and a vision transformer, for 15 epochs with AdamW \citep{adamw} at $10^{-4}$, and report top-1 accuracy over three seeds with 95\% Student-$t$ intervals.
Each dataset is evaluated on the task listed in Table~\ref{tab:datasets}, and all differences in accuracy are given in percentage points of top-1 accuracy.
Because deduplication changes the class distribution of the test set, we also report macro-F1 and balanced accuracy (the mean of the per-class recalls).
These metrics need per-image predictions, which the runs behind Table~\ref{tab:inflation} did not save, so we repeated the ResNet-50 runs on the same \textbf{naive}, \textbf{dedup} and \textbf{ctrl} splits and saved them.
The mean top-1 accuracy of the repeated runs under \textbf{naive} and \textbf{dedup} is within 0.32 points of Table~\ref{tab:inflation} on every dataset.
We also run three shallow baselines that do not learn a representation: LightGBM and logistic regression over each image downsampled to a $32\times32$ grayscale thumbnail and flattened to a 1{,}024-dimensional vector, and 1-NN over its pHash.
We compute the 95\% interval of a drop between two conditions from the half-widths $h_1$ and $h_2$ of their two intervals, as $\sqrt{h_1^2+h_2^2}$.

\begin{table}[t]
\centering
\caption{Top-1 accuracy under the naive, dedup and ctrl splits, mean $\pm$ 95\% CI over three seeds ($^{*}$six, $^{\ddagger}$ten).
Bold drops exclude zero.}
\label{tab:inflation}
\small
\setlength{\tabcolsep}{3.5pt}
\resizebox{\linewidth}{!}{%
\begin{tabular}{lcccc}
\toprule
\textbf{Dataset} & \textbf{naive} & \textbf{dedup} & \textbf{drop} & \textbf{ctrl} \\
\midrule
\multicolumn{5}{l}{\emph{ResNet-50}} \\
Malimg        & 99.61\,\tiny{$\pm$0.08} & 99.11\,\tiny{$\pm$0.56} & $+0.49$\,\tiny{$\pm$0.57} & 98.91\,\tiny{$\pm$0.34} \\
Dumpware10    & 97.98\,\tiny{$\pm$0.88} & 96.69\,\tiny{$\pm$1.13} & $+1.30$\,\tiny{$\pm$1.44} & 96.12\,\tiny{$\pm$3.28} \\
MaleVis       & 96.40\,\tiny{$\pm$1.95} & 90.28\,\tiny{$\pm$1.48} & $\mathbf{+6.12}$\,\tiny{$\pm$2.45} & 85.24\,\tiny{$\pm$1.88} \\
MalNet-Tiny   & 78.73\,\tiny{$\pm$0.42} & 70.28\,\tiny{$\pm$1.89} & $\mathbf{+8.44}$\,\tiny{$\pm$1.93} & 68.31\,\tiny{$\pm$0.57} \\
Virus-MNIST   & 92.30\,\tiny{$\pm$0.97} & 87.05\,\tiny{$\pm$1.78} & $\mathbf{+5.25}$\,\tiny{$\pm$2.03} & 86.17\,\tiny{$\pm$0.89} \\
AndroDex$^{\ddagger}$ & 96.31\,\tiny{$\pm$0.45} & 92.66\,\tiny{$\pm$0.89} & $\mathbf{+3.65}$\,\tiny{$\pm$1.00} & 93.85\,\tiny{$\pm$1.21} \\
MOTIF$^{*}$   & 63.43\,\tiny{$\pm$1.51} & 52.91\,\tiny{$\pm$1.65} & $\mathbf{+10.52}$\,\tiny{$\pm$2.23} & 50.99\,\tiny{$\pm$2.13} \\
\midrule
\multicolumn{5}{l}{\emph{DenseNet-121}} \\
Malimg        & 99.18\,\tiny{$\pm$1.69} & 99.30\,\tiny{$\pm$0.51} & $-0.12$\,\tiny{$\pm$1.77} & --- \\
Dumpware10    & 98.21\,\tiny{$\pm$0.83} & 96.97\,\tiny{$\pm$2.04} & $+1.25$\,\tiny{$\pm$2.20} & --- \\
MaleVis       & 96.97\,\tiny{$\pm$0.90} & 91.46\,\tiny{$\pm$2.35} & $\mathbf{+5.51}$\,\tiny{$\pm$2.51} & --- \\
MalNet-Tiny   & 78.87\,\tiny{$\pm$0.61} & 71.00\,\tiny{$\pm$1.41} & $\mathbf{+7.88}$\,\tiny{$\pm$1.53} & --- \\
Virus-MNIST   & 92.78\,\tiny{$\pm$0.42} & 87.69\,\tiny{$\pm$0.91} & $\mathbf{+5.09}$\,\tiny{$\pm$1.00} & --- \\
AndroDex      & 96.03\,\tiny{$\pm$1.37} & 91.08\,\tiny{$\pm$4.68} & $+4.95$\,\tiny{$\pm$4.88} & --- \\
MOTIF$^{*}$   & 64.81\,\tiny{$\pm$3.61} & 53.38\,\tiny{$\pm$2.96} & $\mathbf{+11.43}$\,\tiny{$\pm$4.67} & --- \\
\midrule
\multicolumn{5}{l}{\emph{ViT-B/16}} \\
Malimg        & 98.84\,\tiny{$\pm$0.94} & 98.23\,\tiny{$\pm$2.46} & $+0.61$\,\tiny{$\pm$2.63} & --- \\
Dumpware10    & 96.31\,\tiny{$\pm$3.67} & 94.55\,\tiny{$\pm$0.41} & $+1.76$\,\tiny{$\pm$3.70} & --- \\
MaleVis       & 95.71\,\tiny{$\pm$0.84} & 89.80\,\tiny{$\pm$0.67} & $\mathbf{+5.91}$\,\tiny{$\pm$1.08} & --- \\
MalNet-Tiny   & 75.28\,\tiny{$\pm$1.48} & 67.00\,\tiny{$\pm$1.63} & $\mathbf{+8.28}$\,\tiny{$\pm$2.20} & --- \\
Virus-MNIST   & 92.41\,\tiny{$\pm$0.92} & 87.10\,\tiny{$\pm$1.03} & $\mathbf{+5.31}$\,\tiny{$\pm$1.38} & --- \\
AndroDex      & 94.07\,\tiny{$\pm$1.57} & 88.76\,\tiny{$\pm$5.05} & $+5.31$\,\tiny{$\pm$5.29} & --- \\
MOTIF$^{*}$   & 58.52 & 47.67 & $+10.85$\,\tiny{$\pm$11.60} & --- \\
\bottomrule
\end{tabular}}
\end{table}

\subsection{Results}
\label{sec:results}

Table~\ref{tab:inflation} reports all three architectures over three seeds, except ResNet-50 on MOTIF and AndroDex, to which we added seeds (six and ten) because the interval on the drop was wide at three.
We run \textbf{ctrl} for ResNet-50 only, over three seeds (six for MOTIF), and take its values from the repeated runs described above.
MalNet-Image, MalVis and MaleX exceed a million images each and are audited but not retrained, and the ViT-B/16 intervals on MOTIF are too wide to be informative (\textbf{naive} alone is $58.52\pm10.92$).

The datasets that lose accuracy are the ones in which the audit found leakage: the two in which it found little or no leakage do not lose a measurable amount (Malimg $+0.49\pm0.57$, Dumpware10 $+1.30\pm1.44$), while the four in which it found substantial leakage lose between 5.3 and 10.5 points.
AndroDex is an intermediate case, since its ResNet-50 interval excludes zero only at ten seeds ($+2.85\pm4.75 \rightarrow +3.65\pm1.00$), while its DenseNet-121 and ViT-B/16 intervals remain at three seeds and still include zero, so the drop is established for one architecture of the three.
On the five datasets that lose accuracy, the shallow baselines lose more than the CNN (6.6--17.5 points for LightGBM, 16.8--33.8 for 1-NN over pHash), so the deep model generalizes beyond texture matching.

\paragraph{Macro-averaged metrics.}
Families differ in how much of their content is duplicated, so deduplication changes the class distribution of the test set.
In seed 0, the share of the largest class in the test set increases from 12.9\% to 29.4\% on MaleVis (the \texttt{Other} class) and from 31.7\% to 45.9\% on Malimg, and families with very few distinct images do not have a test image after deduplication (\texttt{Autorun.K}, \texttt{Obfuscator.AD} and \texttt{Yuner.A} in Malimg, \texttt{VBA} in MaleVis).
The drop in macro-F1 excludes zero on the same five datasets as the drop in top-1 accuracy: it is $6.53\pm2.86$ points on MalNet-Tiny, $15.07\pm1.27$ on Virus-MNIST, $4.31\pm1.33$ on AndroDex, $11.11\pm3.20$ on MOTIF and $13.36\pm11.18$ on MaleVis, while the drops on Malimg ($2.19\pm5.44$) and Dumpware10 ($1.02\pm3.79$) include zero.
Balanced accuracy divides the datasets in the same way, and the per-split values of all three metrics are in our released results.

\paragraph{Leaked and clean test images.}
Within one \textbf{naive} run the model and its training set are fixed, so we measure the effect of leakage without any change in training-set size by comparing the accuracy of the model on the test images that have a near-duplicate in training with its accuracy on the remaining test images.
The leaked test images are classified correctly 94.1--100\% of the time on every dataset.
Because a leaked image may belong to an easier family than a clean one, we reweight the clean accuracy to the family mix of the leaked images, over the families that have both kinds of test image.
After reweighting, accuracy on clean images is lower than on leaked images by $20.75\pm1.16$ points on MalNet-Tiny, $32.94\pm4.48$ on MOTIF, $14.40\pm5.53$ on Virus-MNIST, $20.41\pm12.94$ on MaleVis and $4.78\pm1.75$ on AndroDex, while the differences on Malimg ($5.57\pm6.17$) and Dumpware10 ($2.78\pm5.57$) include zero.

Training on the published training sets of MaleVis, MalNet-Tiny and Dumpware10 and evaluating on their published test sets yields differences of $15.81\pm3.02$, $21.96\pm0.66$ and $1.77\pm1.99$ points, so papers that report results on these splits are affected in the same way.

\paragraph{Training-set size.}
Deduplication reduces the size of the training set, so part of the drop could come from training on fewer images.
The \textbf{ctrl} condition does not separate the two effects, since its training set has fewer distinct images than that of \textbf{dedup} and can contain copies of test images.
We therefore fix the \textbf{dedup} test set of each seed, remove from the pool every near-duplicate group with a member in that test set, and draw two kinds of training set from the rest: one image per group, as in \textbf{dedup}, or random images together with their duplicates.
We train ResNet-50 on each kind at 25, 50, 75 and 100\% of the size of the \textbf{dedup} training set, and also on the whole remaining pool, over three seeds.

The training sets with duplicates are not significantly more accurate at any size, and at 100\% they are less accurate than the training sets of distinct images by $4.73\pm2.13$ points on MaleVis, $2.20\pm0.69$ on MalNet-Tiny and $2.21\pm1.33$ on MOTIF, while Virus-MNIST ($1.12\pm2.05$) and AndroDex ($0.00\pm2.58$) do not show a difference.
The whole remaining pool contains the same groups as the \textbf{dedup} training set and 1.2 to 2.3 times as many images, and training on it yields accuracy within 1.3 points of \textbf{dedup} on all five datasets (MaleVis 90.64 against 90.25, MalNet-Tiny 69.99 against 70.58, Virus-MNIST 86.04 against 86.76, MOTIF 52.16 against 52.39 and AndroDex 93.85 against 92.61).
Accuracy therefore depends on the number of distinct groups in the training set, and adding duplicates of the training images does not raise it toward the \textbf{naive} accuracy, so the drop from \textbf{naive} to \textbf{dedup} comes from test images that have a copy in training and not from the smaller training set.
Our result agrees with that of \citet{dias2026defects} in face recognition, where removing 15\% of the MS1MV3 training images as pHash near-duplicates does not reduce verification accuracy.
Dumpware10 has the smallest deduplicated training set, 2{,}813 images, and its drop ($+1.30\pm1.44$) is not significant.

\paragraph{Temporal split.}
We also split MOTIF by the month of first appearance that its metadata records, training on the samples that appeared up to January 2020 and testing on the 299 samples in 53 families that appeared from February 2020 to May 2021, and compare against random splits with the same number of training and test samples per family.
In the temporal split 2.3\% of the test images have a near-duplicate in training, against 17.4--18.4\% in the random splits, and ResNet-50 is $20.62\pm5.84$\% accurate against $45.37\pm3.84$\% over three seeds.
The temporal split both removes most of the leakage and tests on later malware, so the 25-point difference combines the two effects, and it agrees with the lower accuracy that TESSERACT \citep{tesseract} reports for Android detection under temporal splits.

\paragraph{Relation to published results.}
A drop measured from a weaker training setup would not apply to published results, so we check that our \textbf{naive} accuracy matches each benchmark's published result.
Our \textbf{naive} accuracy exceeds the published result on three of the four benchmarks that report one, Malimg 99.61 against 98.08\% \citep{nataraj2011}, Dumpware10 97.98 against 96.39\% \citep{dumpware10}, and Virus-MNIST 92.30 against roughly 80\% \citep{virusmnist}, and on the fourth, MaleVis, it is within half a point of the published result when measured in the same way, as we describe below.
Three rows do not have a comparable published result: MalNet-Image reports macro-F1 rather than top-1 \citep{malnet}, AndroDex is a data descriptor that does not report a benchmark result of its own \citep{androdex}, and MOTIF-as-byteplot is our rendering.

The 97.48\% that the MaleVis authors report with DenseNet-121 is a closed-set result on their published validation split, which leaves out the legitimate-software class \texttt{Other} \citep{malevis}.
On the 3{,}644 malware images of that split our ResNet-50 yields an accuracy of 97.10\%, or 97.26\% when its prediction is restricted to the 25 malware families.
Including the 1{,}482 \texttt{Other} images lowers top-1 accuracy to 87.66\%, while balanced accuracy is 95.80\%, because the \texttt{Other} images make up 29\% of the published validation images but 4\% of the training images, and 64.5\% of them are classified correctly.
The drops in Table~\ref{tab:inflation} are therefore measured from the reported operating point.

\paragraph{Sensitivity to the near-duplicate threshold.}
The deduplication criterion has one free parameter, the pHash Hamming threshold $\tau$, fixed at 4 throughout.
Figure~\ref{fig:threshold} shows that the measured leakage is stable in $\tau$, and so is the drop: rebuilding the MaleVis splits at $\tau\in\{0,2,8\}$ and retraining yields drops of 3.67, 5.87, 6.12 and 7.34 points at $\tau=0,2,4,8$, monotonic in $\tau$ and never zero.
At $\tau=0$, which collapses only hash-identical images, the drop is still 3.67 points, so $\tau$ sets how much of the drop is measured but not whether there is one.
We quote point estimates because the per-family test composition is fixed across seeds, so other seeds would not measure the variation of this sweep.

\paragraph{Generality across architectures.}
All three architectures lose accuracy on the same datasets, on identical splits.
MaleVis loses 6.12 / 5.51 / 5.91 points and MalNet-Tiny 8.44 / 7.88 / 8.28 under ResNet-50 / DenseNet-121 / ViT-B/16, while Malimg and Dumpware10 remain within noise of zero on all three, so the drop depends on the dataset and not on the architecture.

\paragraph{Shallow baselines.}
On naive splits, LightGBM over the thumbnails is within 5.1 points of the CNN on six of the seven datasets (98.4\% against 99.6\% on Malimg) and 15.6 points below it on MOTIF, while logistic regression is within 5 points only on Malimg and Virus-MNIST.
Deep visual representation learning therefore does not appear necessary for most of the reported accuracies.

\paragraph{Removal rates.}
Reducing each near-duplicate group to one image removes 31.6\% of Malimg, 58\% of MaleVis, 39.4\% of Virus-MNIST and 34.4\% of AndroDex.

\paragraph{Obfuscation.}
AndroDex includes obfuscated variants of its originals, which an obfuscator has rewritten so that their code differs while their behavior is unchanged.
Training on the originals and testing on the obfuscated variants, with zero exact-duplicate leakage between the two, reduces accuracy by 13.6 points (92.76 $\rightarrow$ 79.17), but accuracy remains far above chance, so the byteplot contains some signal that is invariant to obfuscation.
The accuracy that remains is inconsistent with the model relying only on recognizing a fixed, repeated byte pattern such as a packer stub or a shared header.

\section{\uppercase{The Effect of the Image Representation}}
\label{sec:repr}

Section~\ref{sec:inflation} does not establish whether the image encoding is itself responsible for the inflation, because the benchmarks differ in corpus, labeling and rendering at once, so we separate the effect of the rendering from the other two in two ways.

\subsection{Comparing Representations on MOTIF}
The experiment is a paired comparison of two representations of one corpus: we take a fixed set of samples, represent each sample both as a static feature vector and as a byteplot image, train a model on each representation over identical folds, and compare top-1 accuracy.
Because the samples, the folds and the seeds are shared, any difference between the two is attributable to the representation and not to the corpus, the labels or the split.
MOTIF \citep{motif} is the one corpus on which this comparison is possible, since it provides raw binaries and EMBER-v2 \citep{ember} feature vectors for the same samples, the standard static-feature baseline for PE malware.
MOTIF's family labels are also derived from published threat reports rather than from antivirus votes, and its samples are deduplicated by construction.
We render exactly those samples as byteplots and emit fold lists from an identical \texttt{StratifiedKFold} call, so that the image and non-image numbers are paired on samples, folds and seeds.

The MOTIF authors distribute its binaries \emph{disarmed} so that they cannot execute, which has been given as a reason that MOTIF is ``unsuitable as a malware image dataset'' \citep{bruzzese2024}.
Disarming sets two fields of the PE header to zero, the machine type and the subsystem, and restoring them to the values in MOTIF's metadata produces a file with the MD5 hash of the original for 3{,}091 of the 3{,}095 binaries and for all 2{,}964 that we train on.
The rest of the header, the region to which a byteplot CNN attends most \citep{brosolo2025road}, is unchanged, so a byteplot of a disarmed file differs from one of the original file in four pixels.
Retraining ResNet-50 on byteplots of the restored files over the same folds and seeds yields $63.55\pm1.21$, against $63.63\pm0.71$ for the disarmed files (Table~\ref{tab:motif}), so disarming does not affect the comparison with MOTIF's EMBER features, which were computed from the original files.

\begin{table}[t]
\centering
\caption{Representations of the same MOTIF samples, 5-fold CV over seeds 0--2 with identical folds in every row; n/m is not measured.}
\label{tab:motif}
\small
\setlength{\tabcolsep}{3pt}
\begin{tabular}{llrr}
\toprule
\textbf{Model} & \textbf{Representation} & \textbf{top-1} & \textbf{top-5} \\
\midrule
LightGBM      & EMBER-v2 (2{,}381-d)         & $\mathbf{73.93}$\,\tiny{$\pm$0.51} & 82.01 \\
Logistic reg. & EMBER-v2 features             & 73.25\,\tiny{$\pm$0.37} & n/m \\
1-NN          & EMBER-v2 features             & 69.01\,\tiny{$\pm$1.32} & n/m \\
\midrule
ResNet-50     & \textbf{byteplot image}       & $63.63$\,\tiny{$\pm$0.71} & 72.99 \\
LightGBM      & byteplot $32\times32$ pixels  & 48.44\,\tiny{$\pm$0.68} & 58.90 \\
MalConv2      & raw byte sequence             & 48.20\,\tiny{$\pm$0.95} & 59.62 \\
\bottomrule
\end{tabular}
\end{table}

On those folds (Table~\ref{tab:motif}) the byteplot CNN scores $63.63\pm0.71$ against $73.93\pm0.51$ for LightGBM on static features computed from the same binaries, roughly ten points less.
It exceeds the weaker byte-level models, LightGBM on $32\times32$ pixels (48.44) and MalConv \citep{malconv} (48.20), by roughly 15 points, so the byteplot contains usable signal.
MalConv is usually trained on $10^5$--$10^6$ samples, so its accuracy on 2{,}964 is not a ceiling.

MOTIF's feature representation has 0.1\% exact and 2.2\% near-duplicate rows and does not show a drop under deduplication.
Rendering those same samples as byteplots produces 19.7\% near-duplicates (Table~\ref{tab:scorecard}) and a drop of $10.52\pm2.23$ points (Table~\ref{tab:inflation}).
Since the corpus is already deduplicated, the duplication must be introduced by the rendering step, as shared headers, packer stubs and padding all render to near-identical textures.
The rendering is therefore itself a source of duplication, which curation of the corpus cannot prevent.

MaleX \citep{mohammedmalware} allows the same comparison on an independent corpus of 1.04M Windows PE programs, since it provides each program in two encodings: a $256\times256$ byteplot and a bigram-DCT image, the two-dimensional discrete cosine transform of a $256\times256$ histogram of consecutive byte pairs.
Over the same programs, 13.3\% of the byteplots and 6.4\% of the bigram-DCT images are exact content duplicates, and of the 66{,}710 programs whose bigram-DCT image has an exact copy, 93\% also have an exact copy as a byteplot, so the byteplot encoding of this corpus has twice the duplication of the other encoding.
We compare only exact duplicates here, because pHash does not separate families on the bigram-DCT images, where 41.8\% of the pairs at pHash distance 0 have different family labels, against 2.3\% for the byteplots.
MOTIF-as-byteplot is only 63\% accurate and still drops 10 points under deduplication, so a low absolute accuracy does not indicate a clean benchmark.

\subsection{Cross-Rendering Transfer}
The width at which a byteplot is wrapped is an arbitrary choice made by the benchmark's creator, and changing it changes the image without changing any byte of the program.
If a model depends on that choice, then part of what it has learned is a property of the rendering and not of the malware.
Dumpware10 allows a test of this dependence, since it provides the same binaries rendered at four byteplot widths (224, 300, 4096, $\sqrt{\text{filesize}}$).
Re-wrapping a byte stream at a different width is content-preserving, in that no byte changes and only the positions of the row breaks change.
The four renderings of a given binary are near-indistinguishable by eye, sharing global texture statistics to within 1.8 gray levels, yet they differ in 87--90\% of pixels and correlate at only $r=0.44$--$0.51$.
We train one model per width and evaluate each against every width's test split.
Every rendering is resized to $224\times224$, so the width changes where each byte is placed in the image and not the size of the model's input.
Test-set membership is identical across all four renderings (861 binaries; per-binary filename Jaccard $1.000$ in every pair), so each cell varies only in how the same binaries were drawn.

\begin{figure}[t]
\centering
\includegraphics[width=0.88\linewidth]{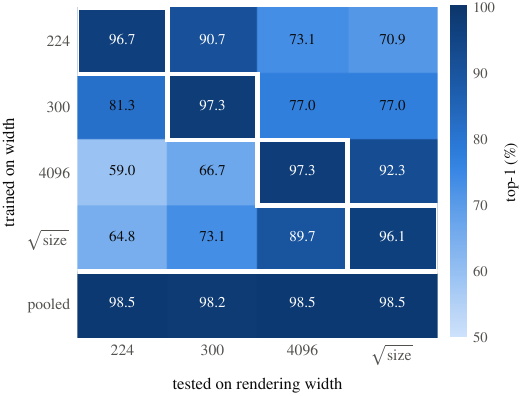}
\caption{Cross-rendering transfer on Dumpware10 (ResNet-50, mean of three seeds).
The outlined diagonal is in-rendering accuracy, and the pooled row is one model trained on all four widths.}
\label{fig:xrender}
\end{figure}

In the transfer matrix of Figure~\ref{fig:xrender}, the mean in-rendering accuracy is $96.84\pm0.48$ and the mean cross-rendering accuracy is $76.31\pm2.84$, a drop of $20.53\pm2.71$ points with a worst cell at 57.07.
A substantial part of what the model uses is therefore the 2-D geometric alignment produced by the reshape.
The structure of the transfer matrix is consistent with a dependence on geometry, since the two narrow widths transfer to each other reasonably well, and 4096 and $\sqrt{\text{filesize}}$ transfer well to each other (89.70, 92.30) but poorly to the narrow widths (59.04--73.05), which suggests that the learned features are specific to a range of widths.

MOTIF is not distributed as images, so we render its binaries ourselves at five widths, the four above and the file-size-dependent width of \citet{nataraj2011} that our MOTIF audit uses, and train ResNet-50 at each width on the first fold of our cross-validation over three seeds.
The mean in-rendering accuracy is $63.78\pm4.69$ and the mean cross-rendering accuracy $30.70\pm3.04$, a drop of $33.08\pm1.74$ points, which is larger than on Dumpware10.

\paragraph{Pooling renderings during training.}
Training a single model on all four renderings pooled yields an accuracy of 98.22--98.49\% on every width, exceeding each width-specialized model on its own width by 1--2 points.
On MOTIF the pooled model is 64.9--65.9\% accurate on every width, 0.1--3.3 points above each width-specialized model.
The dependence on width can therefore be removed in training, and drawing the same binary at several widths is an augmentation that does not change any bytes or require additional data.
We read this observation as one about how benchmarks are built: each benchmark is distributed in a single rendering, so models trained and evaluated on it can depend on that rendering without the dependence being reported.

\citet{brosolo2025road} augment over a transformation that changes content (packing) and recover accuracy, whereas we augment over one that preserves content and both recover transfer and improve accuracy on the diagonal.

\section{\uppercase{Discussion}}
\label{sec:discussion}

The inflation of a benchmark depends on both its corpus and its rendering, so the inflation has to be measured for each benchmark, and we release the tooling for doing so.
Malimg loses only $0.49\pm0.57$ points to content-aware deduplication, an interval that includes zero, and Dumpware10's memory-dump images do not contain any exact duplicates, which makes it the cleanest representation we measured.
A benchmark that passes a deduplication audit can still pose an easy task, as on Malimg, where gradient boosting over $32\times32$ thumbnails is 98.4\% accurate and each of two families consists of a single near-duplicate group.

We make three recommendations based on our measurements.
\textit{(1) Deduplicate on the model's input}, reporting content-hash and perceptual-hash duplication rates alongside accuracy, since MalNet, MalVis and MaleX show that a file-hash deduplication pipeline still leaves 13--17\% of images as content duplicates.
\textit{(2) Render at multiple widths and pool}, which removes the drop under a change of width on both Dumpware10 and MOTIF and improves in-rendering accuracy (Figure~\ref{fig:xrender}), and we do not know of a benchmark that does this.
\textit{(3) Report a non-image baseline on the same samples}, since static features with gradient boosting beat a fine-tuned CNN by 10 points on the one corpus where the comparison is properly paired, so the benefit of an image encoding needs to be demonstrated.

\section{\uppercase{Conclusion}}
\label{sec:conclusion}

In this paper, we audited ten malware-as-image benchmarks for duplication and test-train overlap, rebuilt their splits without the duplicates, and retrained standard classifiers to measure the effect on accuracy.
We find that deduplicating by file hash, which is the standard practice, does not prevent pixel-level leakage, since distinct files render to identical images.
Content-aware splits reduce accuracy by 3.7 to 10.5 points on five benchmarks, while the drop on the other two is not significant.
The rendering introduces duplication that the corpus does not contain, as well as a dependence on wrap width that accounts for a drop of 20 to 33 points.
On MOTIF, where the comparison can be paired, the EMBER-v2 static features outperform the image encoding by 10 points on the same samples.

Duplication of the model's inputs and dependence on the encoding are not specific to malware, since wherever a corpus is converted to a different representation before training, deduplicating the distributed files does not ensure that the model's inputs are distinct, and the encoding can introduce structure that the corpus does not contain.
Duplication of the inputs and dependence on the encoding can both be detected at low cost, by reporting duplication over the model's input and by running one baseline that does not use the encoding, and these two checks would have detected each of the effects we report.

\bibliographystyle{apalike}
{\small
\bibliography{main}}

\end{document}